# General Model of Diffusion of Interstitial Oxygen in Silicon and Germanium Crystals

Vasilii Gusakov

Institute of Solid State and Semiconductor Physics, P. Brovka str. 17, 2200 72 Minsk, Belarus

A theoretical modeling of the oxygen diffusivity in silicon and germanium crystals both at normal and high hydrostatic pressure has been carried out using molecular mechanics, semiempirical and ab initio methods. It was established that the diffusion process of an interstitial oxygen atom ($O_i$) is controlled by the optimum configuration of three silicon (germanium) atoms nearest to $O_i$. The calculated values of the activation energy $\Delta E_a(Si) = 2.59$ eV, $\Delta E_a(Ge) = 2.05$ eV and pre-exponential factor $D_0$ (Si) = 0.28 sm$^2$ s$^{-1}$, $D_0$ (Ge) = 0.39 sm$^2$ s$^{-1}$ are in a good agreement with experimental ones and for the first time describe perfectly an experimental temperature dependence of the $O_i$ diffusion constant in Si crystals (T=350 - 1200°C). Hydrostatic pressure (P≤80 kbar) results in a linear decrease of the diffusion barrier ($\partial_P \Delta E_a(P) = -4.38 \cdot 10^{-3}$ eV kbar$^{-1}$ for Si crystals). The calculated pressure dependence of $O_i$ diffusivity in silicon crystals agrees well with the pressure enhanced initial growth of oxygen-related thermal donors.

Development of theoretical methods of determining the diffusivity of atoms in crystals is of great interest not only from fundamental, but also from practical point of view. The reasoning is that the atomic diffusion in crystals occurs very often under extreme conditions (very high temperatures, fields of stress etc) and that essentially impedes, makes expensive or even impossible an experimental research. However till now there are many obscure questions relating to the microscopic mechanism of diffusion in crystals, whenever migration of an impurity atom involves the braking and forming of covalent bonds. It is common knowledge the diffusion of interstitial oxygen atoms in silicon crystals is of crucial importance in the processes of oxygen agglomeration (formation of thermal donors) and in the gathering of metallic impurities in industrial processing of silicon and, as a result, the experimental measurements of the diffusivity of oxygen in silicon has received much attention. As pointed out by Mikekelsen [1] most experimental data can be consistently fit over a wide temperature range (350 – 1200 K) by a single expression of the form $D = 0.13 \exp(-2.53 eV / k_B T)$ sm$^2$ s$^{-1}$. The expression has been obtained by fitting to data from six independent experiments. This expression is generally believed to be the intrinsic diffusion constant involving oxygen jumping from a bond-center to one of the six nearest bon-center cites.

Several theoretical efforts have attempted to calculate the diffusion barrier, but different results were obtained. Thus the calculated values of the barrier are ranging from 1.2 eV [2], 2.0 eV [3] up to 2.3 eV [4], 2.5 eV [5]. All these calculations (except [6]) assume the saddle point configuration for diffusion in a (110) plane and the midway between the two bond-center sites. The remaining degrees of freedom and the position of the other Si atoms were determined by total-energy minimization. The resulting total energy, measured from the energy of the equilibrium configuration, results the adiabatic activation energy for diffusion. Using empirical interatomic potentials Jiang and Brown[6] have concluded that the saddle point of $O_i$ migration is past the midpoint, but their conclusion has called in question in [7]. Moreover Ramamoorthy and Pantelides[7] have offered that a seemingly simple oxygen jump is actually a complex process can be properly described in terms of coupled barriers by energy hypersurface with an activation energy is ranging from 2.2 eV up to 2.7 eV. In this connection it should be pointed out that the calculation of an activation barrier is important, but not an ultimate point of the theoretical description of a diffusion constant. The complete calculation of the diffusion coefficient guesses calculation and the pre-exponential factor. Unfortunately in the majority of previously published works the pre-exponential factor was not evaluated at all, and the calculated value in [6] differs from the experimental one more than on an order of magnitude.

In this Letter, the simulation of diffusion of interstitial oxygen (Oi) in silicon and germanium crystals under normal and hydrostatic pressure (HP) is reported. The activation barrier and pre-exponential factor have been calculated and are in excellent agreement with experimental ones. To the best of my knowledge, no effects of HP on the Oi diffusivity have been considered yet.

Let us consider the physical parameters determining the diffusion process of an atom in a crystal. Modeling by the method of casual wanderings results in the following general expression for the diffusion constant:

$$D = \frac{d^2 N_{et}}{2 \dim} \Gamma, \qquad (1)$$

where $d$ is diffusion jump distance, $N_{et}$ is the number of the equivalent trajectories leaving the starting point, $dim$ is the dimension of space, $\Gamma$ is the average frequency of jumps on the distance $d$. In the case of a system consisting of $N$ atoms, using the reaction-rate theory [8], the value of $\Gamma$ may be written in the following form

$$\Gamma = \frac{1}{2\pi} \frac{\prod_{i=0}^{N} \lambda_i^{(o)}}{\prod_{i=1}^{N} \lambda_i^{(b)}} \exp\left(-\frac{\Delta E_a}{k_B T}\right), \qquad (2)$$

where $\Delta E_a$ is the adiabatic potential energy difference between the saddle point and the stable one, $\lambda_i^{2(o,b)}$ are the eigenvalues of the matrix (with respect to mass-weighted internal coordinates) $K_{ij} = \partial^2 U_{eff} / \partial f_i \partial f_j$, $U_{eff}(f_1,...,f_m)$ denotes the potential function as a function of the internal degrees of freedom. The indices (b) and (o) indicate that the corresponding quantities are evaluated at the saddle point and local minimum, respectively. Thus, the diffusion constant $D$ is determined by the following diffusion parameters: the length of diffusion jumps ($d$), the diffusion barrier ($\Delta E_a$), the number of equivalent ways leaving the starting point of diffusion jumps ($N_{et}$) and the eigenvalues matrix ($\lambda_i^{2(o,b)}$). The calculation of diffusion parameters was performed in a cluster approximation. For comparison with the previous calculations different methods such as empirical potential (MM2), semiempirical (AM1, PM3, PM5) and ab initio (RHF, LDA) have been used for the calculation of the cluster total energy. Depending on the method of total energy calculation the cluster size was varied from 17 Si atoms (ab initio methods) up to $10^3$ Si atoms (semiempirical and empirical potential methods).

Individual oxygen atoms occupy interstitial bond-center (BC) position in silicon and to diffuse by jumping between the neighboring BC sites. Hence the starting and the final points of the diffusion jump correspond to the equilibrium configuration of an interstitial oxygen atom ($O_i$) in silicon. The calculated equilibrium configuration of $O_i$ and the local vibration frequency of the asymmetric stretching mode ($B_1$) takes the following values: $d_{Si-O}$=1.63 Å (6-31G$^{**}$), 1.582 Å

(MM2), 1.61 Å (PM5), $\angle_{Si-O-Si}$=161.6° (6-31G**), 167.3° (MM2), 171° (PM5), $\nu_O$= 1214 cm$^{-1}$ (6-31G**), 1091 cm$^{-1}$ (AM1), 1078 cm$^{-1}$ (LDA) and are in a good agreement with experimental [9,10] and recently calculated ones [11,12]. The calculated value of the potential barrier for the rotation of O$_i$ around Si-Si axis equals $\Delta E_\varphi \leq 20$ meV (PM5). As $\Delta E_\varphi$ is much less than $k_B T$ (at diffusion temperatures) an interstitial oxygen atom may jump on any of six nearest Si-Si bonds and, hence, in Formula (1) the parameters $N_{ef}$= 6 and $d$=1.9 Å.

In the course of transition of O$_i$ atom from one equilibrium configuration in another the braking of old and formation of new covalent Si-O bonds takes place. The process of reconfiguration of an electronic subsystem will occur in the case when oxygen and neighboring silicon atoms owing to thermal fluctuations get in the region of configuration space G (bounded by the critical surface $S_G$) where the electronic reconfiguration leads to lowering of the crystal total energy. It is clear that for the given position of the oxygen atom there are many configurations of silicon atoms for which the electronic reconfiguration can occur but all of these configurations are differ in the total energy of a crystal. Since the diffuse constant exponentially depends on the diffusion barrier (2) we should select the minimal value of $\Delta E_a$

$$\Delta E_a = \inf[E_{cl}(S_G) - E_{cl}(O)], \quad (3)$$

where $E_{cl}(S_G)$ and $\Delta E_{cl}(O)$ are the total cluster energy on the surface $S_G$ and in local minimum O (equilibrium O$_i$ configuration), respectively. In our simulation the value of $\Delta E_a$ was calculated as follows. The oxygen atom was displaced from the equilibrium O$_i$ configuration along a trajectory in the direction of the nearest Si-Si bond. Along the given trajectory the total cluster energy has been calculated and among the set of calculated trajectories the extreme trajectory satisfying condition (3) was selected. It is significant that for the extreme trajectory (3) the saddle point of O migration is displaced from the midpoint of the path both for Si and Ge crystals and the displacement is far more for Ge crystals. The simulation has revealed an important fact for understanding of the diffusion process. $E_{cl}(S_G)$ and hence $\Delta E_a$ depends on the number of the nearest to O$_i$ silicon atoms ($n$) involved in the minimization of the cluster total energy. The diffusion barrier $\Delta E_a(n)$ decreases and tends to 2 eV with increase of the number of Si atoms involved in minimization (Fig 1.). Therefore, first of all, we should determine how many of the nearest to O$_i$ silicon atoms are involved in the diffusion process. An O$_i$ atom can overcome a barrier at any optimum configuration of the nearest Si atoms. However the relative number of $O_i$ atoms diffused at the given optimum configuration is proportional to the product of probabilities of occurrence of the optimum configuration ($P_{occ}$) and probability of diffusion jump ($P_{dj} \propto \exp(-\Delta E_a(n)/k_B T)$). The probability of occurrence of an optimum configuration out of n atoms have been calculated on the basis of geometrical definition of probability (a problem of random collisions) and in this case the product $P_{occ}P_{dj}$ may be written as

$$P(n) \equiv P_{occ}P_{dj} \propto n\left(\frac{\Delta\tau(n)}{\tau(n)}\right)^{n-1} \times$$
$$\times \exp\left(-\frac{\Delta E_a(n)}{k_B T}\right), \quad (4)$$

where $n$ is the number of atoms in the optimum configuration, $\tau(n)$ and $\Delta\tau(n)$ are the period of formation and lifetime of the given optimum configuration, respectively, $\Delta E_{ai}(n)$ is the diffusion barrier. Usually $\Delta\tau(n)/\tau(n)$ is much less than one [8]. The dependence $P(n)$ as function of Si atoms involved in the minimization is depicted in the inset of Fig. 1. Calculations have shown, that $P(n)$ has a sharp maximum at $n=3$ which more than on the order of magnitude exceeds $P(n)$ for $n=2, 4, 5...$ Hence, essentially all $O_i$ atoms overcome the diffusion barrier when only *three* nearest Si atoms are in the optimum configuration, and the diffusion parameters should be calculated for the given configuration. For this case the following values of the diffusion barrier $\Delta E_a = 2.59 - 2.6$ eV (AM1, PM3, PM5 method of calculation) have been obtained. Matrix $\lambda_i^{2(o,b)}$ necessary for the calculation of the pre-exponential factor $D_0$ was evaluated as follows. At the equilibrium configuration of interstitial oxygen $O_i$ and at the intersection point of the extreme trajectory of $O_i$ with surface $S_G$ the square-law interpolation of the potential energy $U_{eff}(f_1,...,f_m)$ ($f_1...f_m$ are coordinates of $O_i$ and nearest Si atoms) has been constructed and $\lambda_i^{2(o,b)}$ has been obtained at once by diagonalization of $K_{ij}$. In such a manner calculated value of the pre-exponential factor equals $D_0 = 0.28 - 0.3$ sm$^2$ s$^{-1}$. On Fig. 2 one can see the excellent agreement between the calculated and experimental temperature dependences of the diffusion coefficient in all temperature range T=350 – 1200 K.

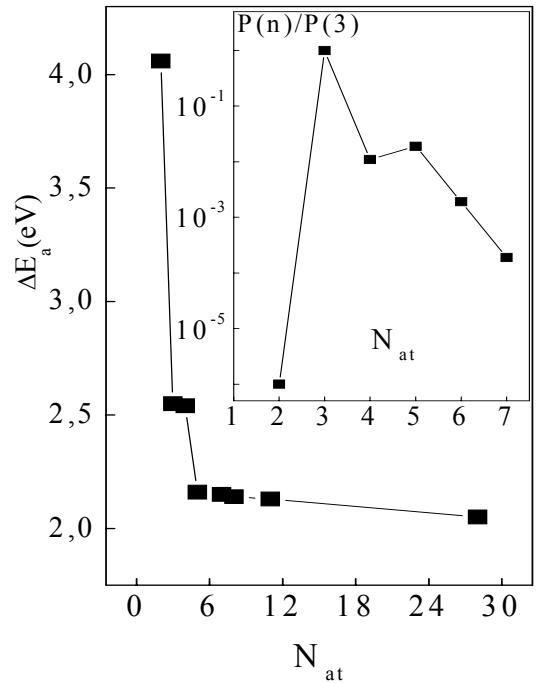

**Fig. 1. Diffusion barrier $\Delta E_a(n)$ as a function of the number of Si atoms involved in minimization of the total cluster energy. In the inset the probability $P_{occ}P_{dj}$ of occurrence of an optimum configuration out of n atoms is presented ($\Delta \tau(n)/\tau(n)$=0.01).**

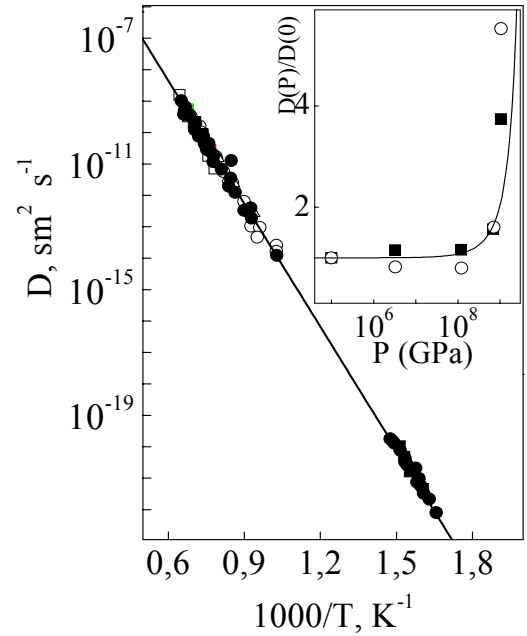

**Fig. 2. Temperature dependence of diffusion constant of interstitial oxygen atom in silicon. Points – experiment [1], line – theory. In the inset: solid line indicates the calculated pressure dependence of a relative coefficient of diffusion $D(P)/D(0)$, T= 450ºC; points are the relative concentration of oxygen thermal donors as a function of pressure (experiment [16]).**

Being grounded on the procedure described above, the diffusion coefficient of interstitial oxygen in a germanium crystals have been calculated also. Calculated values of the activation energy $\Delta E_a$ (Ge) = 2.05 eV and pre-exponential factor $D_0$ (Ge) = 0.39 sm$^2$ s$^{-1}$ are in excellent agreement with experimental ones

$\Delta E_{exp}$(Ge) = 2.076 eV, $D_{exp}$ (Ge) = 0.4 sm$^2$ s$^{-1}$ [13].

For better understanding of oxygen diffusion in silicon crystals the influence of hydrostatic pressure (*P*) on the diffusion coefficient has been evaluated. This is particularly interesting as the high hydrostatic pressure (HP) has been found to enhance strongly the oxygen agglomeration at elevated temperatures[14,15,16,17]. The origin of this unusual phenomenon has been under debate and remains open. To explain the HP effect in [15,16] an enhancement and in [17] an opposite effect of retardation of oxygen diffusion occurred at high temperatures under HP has been suggested. In experimental studies of agglomeration processes of oxygen in silicon hydrostatic pressure usually reaches 1-1.5 GPa. At given pressures variation of Si lattice constant is relatively small ($\leq$ 0.1 Å) and, hence, changes of diffusion coefficient will be determined by variation of $\Delta E_a$ ($D \propto \exp(-\Delta E_a / kT)$) with pressure. To model the effect of pressure the cluster has been conventionally divided into internal ($R < R_0$) and external parts ($R > R_0$). The internal part includes $O_i$ and is selected in such a manner that the increase in $R_0$ does not result in essential change of the equilibrium structure of $O_i$ defect (Si-O bonds and Si-O-Si angle) at *P*=0 (usually $R_o$ equals 5-7 Å). The pressure has been modeled by replacement of the equilibrium length of Si-Si bonds in the external part of the cluster with the length of Si-Si bonds that are characteristic (calculated from experimental value of Si compressibility modulus) for the given pressure. Upon minimization of the cluster total energy, the lengths of Si-Si bonds at $R > R_0$ did not vary, and the minimization was carried out on the coordinates of oxygen and silicon atoms being in the internal part of the cluster. The further evaluation of $\Delta E_a(P)$ was carried out similarly to the case *P*=0. Calculations have revealed that hydrostatic pressure leads to a lowering the diffusion barrier $\Delta E_a(P)$ and in the whole investigated interval of pressures ($P \leq 80$ *kbar*) is described well by the following expression:

$$\Delta E_a(P) / \Delta E_a(0) = 1 - \gamma P, \quad (5)$$

where $\gamma$ = 1.69 10$^{-3}$ kbar$^{-1}$, *P* is the hydrostatic pressure in kbar. The calculated pressure dependence of the $O_i$ diffusivity (without any adjustable parameters) corresponds well to an enhanced growth of the oxygen-related thermal donors (TDs) observed experimentally [15]. Fig. 2 shows the calculated diffusion constant of an interstitial oxygen atom in silicon and the experimentally observed dependence of relative concentration of TD as a function of pressure. One can see that the theoretical curve is well consistent with a sharp increase in the TD enhanced growth at *P* ~ 1GPa.

In summary, a theoretical modeling of the oxygen diffusivity in silicon and germanium crystals at normal and uniform pressure has been presented. On the basis of the results obtained it is possible to draw the following conclusions. Three nearest Si (Ge) atoms are involved in an elementary oxygen jump from a bond-center site to another bond-center site along a path in the (110) plane. It is precisely their optimum position (corresponding to a local minimum of the crystal total energy) determines the value of the diffusion potential barrier of an interstitial oxygen atom in silicon and germanium. The theoretically determined values of the diffusion potential barrier and pre-exponential factor are in excellent agreement with experimental ones and describe very well the experimental temperature dependence of diffusion constant in Si crystals (T = 350 - 1200°C). Hydrostatic pressure ($P \leq 80$ *kbar*) gives rise to the decrease of the diffusion potential barrier in Si crystals and accordingly increases the diffusion coefficient. Such a pressure dependence of $O_i$ diffusivity appears most likely to be responsible for the HP enhancement in generation of the oxygen-related thermal donors.

Financial support by CADRES and INTAS 03-50-4529 is acknowledged.